\documentclass[twocolumn,tighten]{aastex631}

\usepackage{amsmath}
\definecolor{mygreen}{rgb}{0.19,0.55,0.11}
\usepackage{natbib}

\shorttitle{Evolution of NS-He WD bianries}
\shortauthors{Chen et al.}

\graphicspath{{./}{figures/}}

\begin{document}

\title{Investigating the stability of mass transfer in neutron star--helium white dwarf binaries}

\correspondingauthor{Hai-Liang Chen}
\email{chenhl@ynao.ac.cn}

\author{Hai-Liang Chen}
\affiliation{Yunnan Observatories, Chinese Academy of Sciences (CAS), Kunming 650216, P.R. China}

\author[0000-0002-3865-7265]{Thomas M. Tauris}
\affiliation{Department of Materials and Production, Aalborg University, Skjernvej 4A, DK-9220~Aalborg {\O}st, Denmark}

\author{Xuefei Chen}
\affiliation{Yunnan Observatories, Chinese Academy of Sciences (CAS), Kunming 650216, P.R. China}
\affiliation{University of the Chinese Academy of Sciences, Yuquan Road 19, Shijingshan Block, 100049, Beijing, China}

\author[0000-0001-9204-7778]{Zhanwen Han}
\affiliation{Yunnan Observatories, Chinese Academy of Sciences (CAS), Kunming 650216, P.R. China}
\affiliation{University of the Chinese Academy of Sciences, Yuquan Road 19, Shijingshan Block, 100049, Beijing, China}

\begin{abstract}
Neutron star--helium white dwarf (NS+He~WD) binaries are important evolutionary products of close-orbit binary star systems.
They are often observed as millisecond pulsars and may continue evolving into ultra-compact X-ray binaries (UCXBs) 
and continuous gravitational wave (GW) sources that will be detected by space-borne GW observatories, such as LISA, TianQin and Taiji. 
Nevertheless, the stability of NS+He~WD binaries undergoing mass transfer is not well studied and still under debate. 
In this paper, we model the evolution of NS+He~WD binaries with WD masses ranging from $0.17 - 0.45\;M_{\odot}$, applying the detailed stellar evolution code \textsc{mesa}. Contrary to previous studies based on hydrodynamics, we find that apparently {\em all} NS+He~WD binaries undergo stable mass transfer. 
We find for such UCXBs that the larger the WD mass, the larger the maximum mass-transfer rate and the smaller the minimum orbital period during their evolution. Finally, we demonstrate numerically and analytically that there is a tight correlation between WD mass and GW frequency for UCXBs, independent of NS mass.
\end{abstract}


\section{Introduction}\label{sec:intro}
Among the common products of X-ray binaries are neutron star---white dwarf (NS+WD) systems which are important for studies of close binary evolution \citep[e.g.][]{tv23}. Due to gravitational wave (GW) radiation in NS+WD binaries with short orbital periods, the WD may overfill its Roche lobe and the systems become semi-detached within the age of the Universe. If their mass transfer is dynamically stable, they evolve into long-lived ultra-compact X-ray binaries \citep[UXCBs,][]{nrj86}. Because of their short orbital periods, their GW signals can be detected by space-borne low-frequency GW observatories like LISA \citep{aabb+17}, TianQin \citep{lcdg+16}, and Taiji \citep{rgcz20}. If their mass transfer is unstable, the systems eventually merge and may give rise to transient events \citep[e.g.][]{zpt19,zbp20,fmm19,bzpd+22}, producing exotic objects such as Thorne-\.Zytkow-like objects \citep{ples11}, long gamma-ray burst \citep{kod07} or fast radio burst \citep{katz21}.

The stability of mass transfer in NS+He~WD binaries, however, is still a topic of debate and could strongly influence the predicted number of UCXBs as GW sources.
In earlier studies, it was common to use a critical WD mass, or a threshold mass ratio between the WD and the NS, to predict the stability of the mass transfer. Some studies \citep[e.g.][]{vr88,pmbs09,ynv02,vnvk12,ylj21} adopted semi-analytic methods to model the evolution of NS+WD binaries and showed that the critical WD mass is approximately $0.37-1.25\;M_{\odot}$, mainly depending on the stability criteria of mass transfer. However, in these studies, the detailed structure of the WD was not considered and the WD was assumed to be completely degenerate. Given that He~WDs usually have an outer layer of non-degenerate hydrogen, possibly still relatively hot at the moment mass transfer is initiated, such semi-analytic studies may not be realistic, as demonstrated by e.g. \citet{itla14,imtl+16,stli17,taur18}.

\citet{bdc17} carried out hydrodynamic simulations of mass transfer for NS+WD binaries. They found that disk winds appear during the mass-transfer process and measured the specific angular momentum loss of the disk winds. By combining the results of hydrodynamic simulations with a semi-analytic long-tem evolution model, they found that the critical WD mass is only $0.20\;M_{\odot}$, i.e. much smaller than the results from other studies. They also found that the critical WD mass strongly depends on the angular momentum loss from the binary system.

Motivated by the work of \citet{stli17} and \citet{cthc21}, who modelled the detailed structure of He~WDs in UCXBs (produced as NS companion remnants in low-mass X-ray binaries, LMXBs), we model here the evolution of mass transfer in NS+He~WD binaries with a range of different WD masses using the stellar evolution code Modules for Experiments in Stellar Astrophysics (\textsc{mesa}, \citealt{pbdh+11,pcab+13,pmsb+15,psbb+18,pssg+19}). 

The remainder of this paper is organized as follows. In Section~\ref{sec:met}, we introduce the code and the assumptions we adopt for our simulations. In Section~\ref{sec:res}, we present the results we obtained. In Section~\ref{sec:dis}, we discuss uncertainties and their influence on our simulations. In addition, we discuss general properties of UCXBs as GW sources. Finally, we summarize our conclusion in Section~\ref{sec:con}.

\section{Method and assumptions}\label{sec:met}
\subsection{Initial He~WD models}
It is known that NS+He~WD binaries in the Galactic disk can be produced via either: i) stable mass transfer (Roche-lobe overflow, RLO) onto a NS from a main-sequence or a red-giant star companion \citep[e.g.][]{ts99,itl14,cthc21}, or ii) through common envelope (CE) evolution that ejects the hydrogen-rich envelope of the companion star \citep[e.g.][]{ty93,ijcd+13}. In addition, in globular clusters, they can also form via exchange encounter events \citep[e.g.][]{ihrb+08}. The NS+He~WD systems that are produced via stable RLO, and evolve to become semi-detached UCXBs within the age of the Universe, have He~WD masses $<0.17\;M_{\odot}$ \citep{stli17,taur18,cthc21}. UCXBs with larger initial WD masses can only be produced via the CE channel, or in an exchange encounter event in a dense stellar environment. 

Because it is difficult to model in detail the outcome of a CE process or an exchange encounter event, we apply here a method where we first produce a range of He~WD models with different masses between $0.17-0.45\;M_\odot$ via stable RLO in LMXBs with different initial orbital periods (see below), and then we artificially pair these He~WDs in new tight orbits with a NS such that the WDs will be forced to fill their Roche lobe and initiate mass transfer. This approach allows us to study the stability of the RLO in NS+He~WD (UCXB) systems.

We start by computing the evolution of a grid of LMXBs with the \textsc{mesa} code following the method described by \citet{itl14}. The initial donor in this grid is assumed to be a ZAMS star with a mass of $1.2\;M_{\odot}$. The initial orbital period ranges between $1.0 - 600\;{\rm days}$. The initial NS mass is assumed to be $1.30\;M_{\odot}$ (but is irrelevant here in the LMXB stage that is only needed to produce WD models for the next UCXB stage).
After the LMXB stage, we extract and save the He~WD model when its central temperature is $\sim 10^{7}\;{\rm K}$.
The WD masses and their central and surface temperatures are listed in Table~\ref{tab:ini_wd}.                                                                                                                                    

Afterwards, in the second stage, we take these WD models as the initial WD models of NS+He~WD binaries in our UCXB simulations. Compared with NS+He~WD binaries produced from CE evolution, the He~WDs may have different temperatures and residual hydrogen envelope masses. However, as we shall see, these factors have little impact on our results.

\subsection{Evolutionary models of NS+He~WD binaries}
To obtain realistic mass-transfer rate evolution for NS+He~WD binaries, we evolve a NS accretor and a He WD companion with the \textsf{star\_plus\_point\_mass} test suite of \textsc{mesa} code (version 12115). We thus take the NS as a point mass and assume its mass to be $1.30\;M_{\odot}$. The initial orbital periods and WD properties, as well as those at the subsequent onset of RLO, are listed in Table~\ref{tab:ini_wd}. To secure approximately similar central WD temperatures in all our models (and to help the code to smoothly converge), the initial orbital periods are chosen so that the He~WDs do not become too cold before they overfill their Roche-lobe radii. 

\begin{table*}[]
    \centering
    \begin{tabular}{|c|c|c|c||c|c|c|c|c|}
    \hline
    $M_2\;(M_\odot)$ & $P_{\rm orb}\;({\rm days})$ & $\log_{10}(T_{\rm c}/{\rm K})$ & $\log_{10}(T_{\rm eff}/{\rm K})$ & $P^{\rm RLO}_{\rm orb}\;({\rm days})$ & $\log_{10}(T^{\rm RLO}_{\rm c}/{\rm K})$ & $\log_{10}(T^{\rm RLO}_{\rm eff}/{\rm K})$ & $P_{\rm min}\;({\rm min})$ \\
    \hline
    0.17 &  0.30 & 6.995 & 3.88 & 0.0051 & 6.898 & 3.79 & 4.79\\
    0.21 &  0.20 & 6.992 & 3.93 & 0.0038 & 6.934 & 3.88 & 3.78\\ 
    0.25 &  0.05 & 6.996 & 3.96 & 0.0029 & 6.994 & 3.96 & 3.08\\ 
    0.30 &  0.05 & 6.991 & 3.98 & 0.0022 & 6.989 & 3.98 & 2.45\\ 
    0.35 &  0.02 & 6.998 & 4.01 & 0.0017 & 6.997 & 4.01 & 2.04\\ 
    0.40 &  0.02 & 6.994 & 4.02 & 0.0014 & 6.994 & 4.01 & 1.72\\ 
    0.43 &  0.02 & 6.997 & 4.03 & 0.0013 & 6.997 & 4.03 & 1.56\\ 
    0.45 &  0.02 & 6.997 & 4.03 & 0.0012 & 6.997 & 4.03 & 1.48\\ 
    \hline
    \end{tabular}
    \caption{Parameters of simulated NS+He~WD binaries. The first column is the initial WD mass; columns two to four are the initial model values of orbital period, central temperature, and effective temperature; columns five to seven are the same values at the onset of RLO; the eighth column is the minimum orbital periods during their UCXB evolution.}
    \label{tab:ini_wd}
\end{table*}

In our calculation, two kinds of angular momentum loss mechanisms are considered: GW radiation and angular momentum loss due to mass loss. 
The following formula is adopted to compute the angular momentum loss due to GW radiation \citep{ll71}:
\begin{equation}
	\frac{{\rm d}J_{\rm gw}}{{\rm d}t} = -\frac{32}{5}\frac{G^{7/2}}{c^{5}}\frac{M^2_{\rm NS}M^2_2 (M_{\rm NS}+M_2)^{1/2}}{a^{7/2}}\;,
\end{equation}
where $G$ and $c$ are the gravitational constant and the speed of light in vacuum, respectively; $a$ is the binary separation; $M_{\rm NS}$ and $M_2$ are the masses of the NS and the He WD, respectively. 

In our work, the \textsf{Ritter} scheme is adopted to compute the mass-transfer rate \citep{ritt90}. We have tested that the mass-transfer scheme has no significant influence on our results. Moreover, the isotropic re-emission model of mass transfer is also adopted \citep{tv06}. In this model, we assume that the material is transferred conservatively to the NS, and a fraction of this material, $\beta$ is lost from the NS, taking away the specific orbital angular momentum of the NS. Here we adopt an accretion efficiency of 30\% for the NS, i.e. $\beta =0.70$ \citep{ts99,avkf+12}. 
Our results are basically independent of the NS accretion efficiency (Section~\ref{subsec:acc_efficiency}), and also independent of the initial NS mass in the UCXB system (Section~\ref{subsec:initial_NS_mass}).

In addition, we assume that the accretion of the NS is limited by the Eddington mass-accretion rate given by the following expression:
\begin{equation}
	\dot{M}_{\rm Edd} = \frac{4\pi\,GM_{\rm NS}}{\eta \,0.20\,(1+X) c}\;,
\end{equation}
where $X$ is the hydrogen mass fraction of the accreted material. 
After the hydrogen shell of the He~WD is removed, $X=0$.
We assume the ratio of gravitational mass to baryonic mass of accreted material to be 0.85, i.e. $\eta = 0.15$ \citep[e.g.][]{lp07}. The inlist files for our simulations will be available on the \dataset[Zenodo communities]{https://zenodo.org/communities/mesa?page=1&size=20}.

\section{Results}\label{sec:res}
\subsection{An example of UCXB evolution}\label{subsec:exam}

\begin{figure}
    \centering
    \includegraphics[width=\columnwidth]{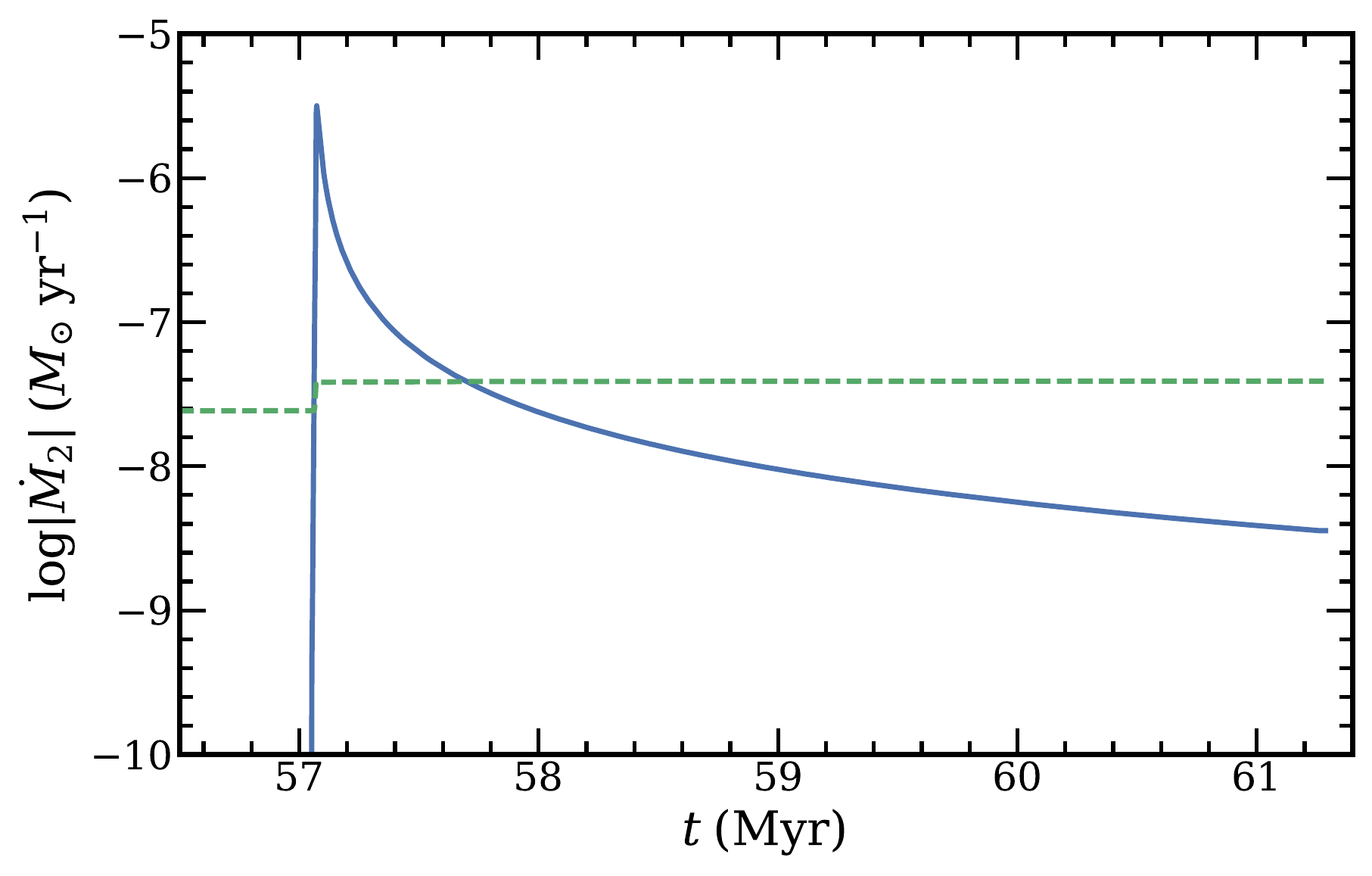}
    \includegraphics[width=\columnwidth]{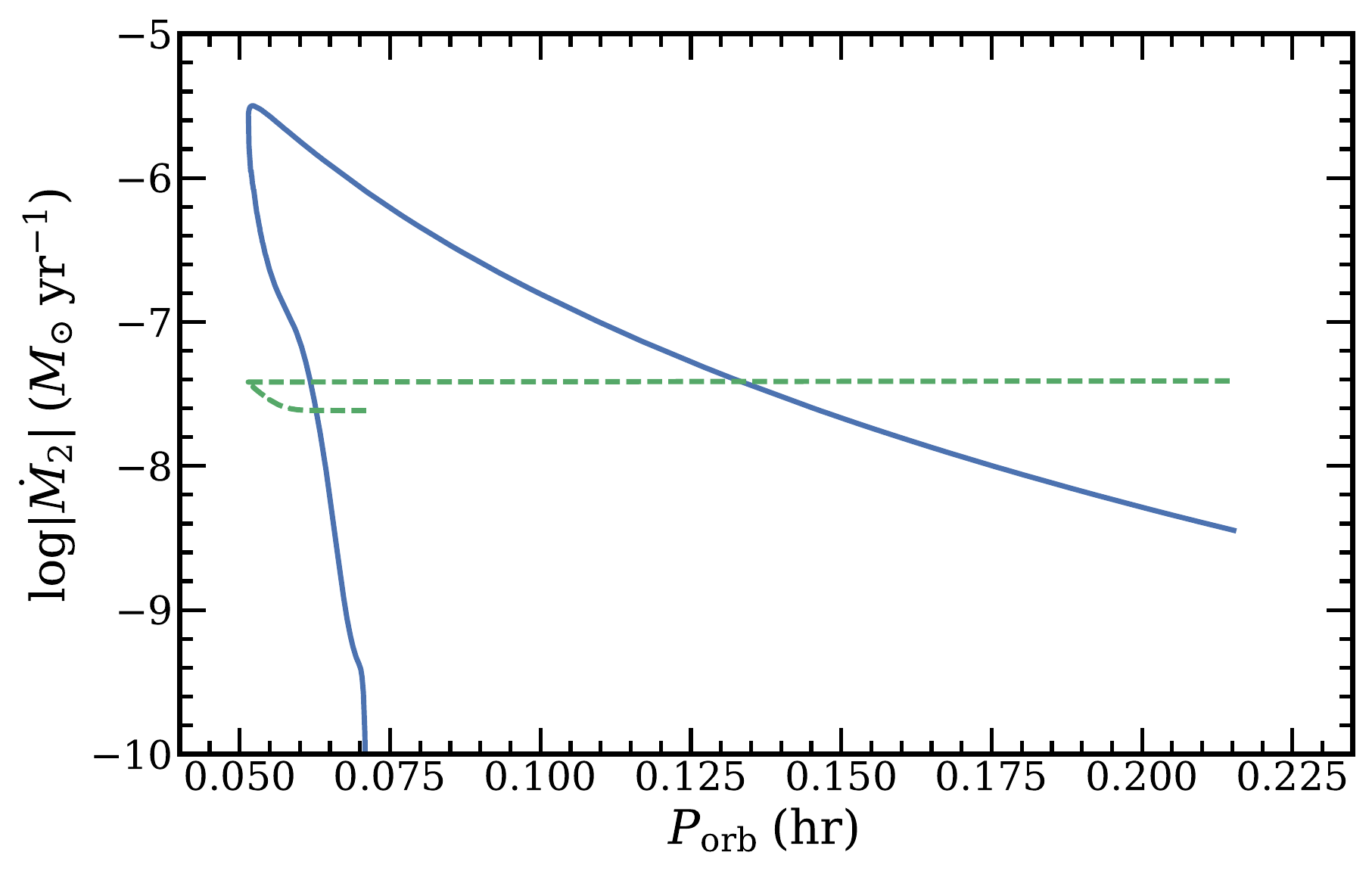}
    \includegraphics[width=\columnwidth]{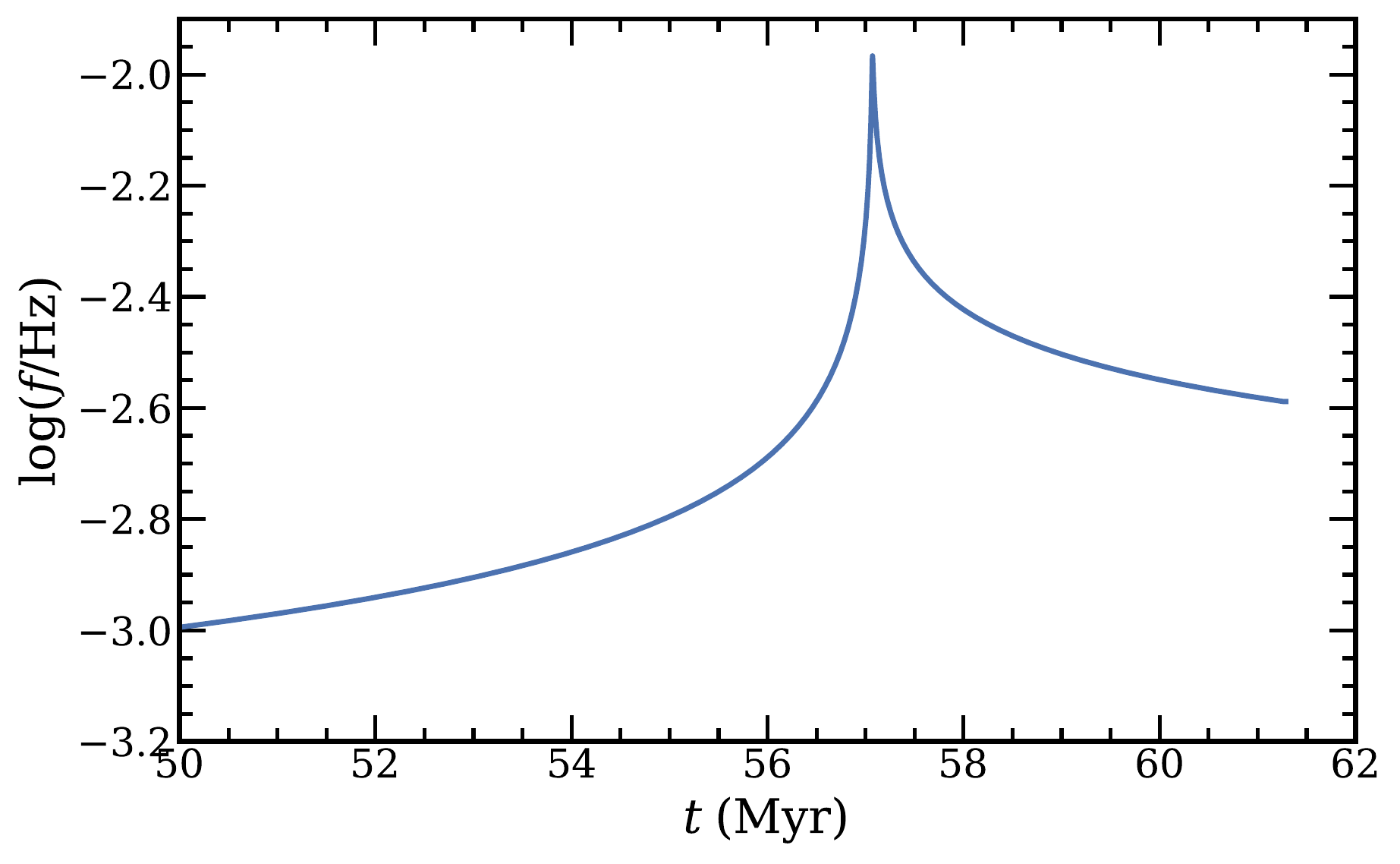}
    \caption{Example of evolution of a NS+He~WD UCXB. The initial binary parameters in this example are $M_{2} = 0.25\;M_{\odot}$, $P_{\rm orb} = 0.05\;{\rm days}$. The upper and middle panels show the evolution of mass-transfer rate as a function of time and orbital period, respectively. The lower panel shows the evolution of GW frequency as a function of time. 
   The dashed lines in the upper two panels indicate the Eddington accretion rate (which is not a constant due to the decrease in hydrogen in of the outer layer of the mass-loosing WD). }
    \label{fig:bin_ev_ex}
\end{figure}

In Fig.~\ref{fig:bin_ev_ex}, we show an example of UCXB evolution of a NS+He~WD system. The initial WD mass in this case is $0.25\;M_{\odot}$ and the initial orbital period is $0.05\;{\rm days}$. During the initial detached phase, the binary separation decreases because of GW radiation and thus the GW frequency increases. At $t \simeq 5.7 \times 10^{7}\;{\rm yr}$, the WD fills its Roche lobe and initiates mass transfer.
At the early stage of RLO, the orbital angular momentum loss due to GW radiation continues to dominate, leading to further increase of the GW frequency. When the orbital period reaches its minimum ($P_{\rm min}\sim 3.08\;$min), the WD has lost its outer non-degenerate layer and the mass-transfer rate is close to its maximum value. 
Hereafter, the rate of orbital expansion dominates over the orbital decay due to GWs and the mass-transfer rate decreases with time. Consequently, the GW frequency also decreases as the orbit widens.

\subsection{NS+He~WD binaries with different WD masses}
\begin{figure}
    \centering
    \includegraphics[width=\columnwidth]{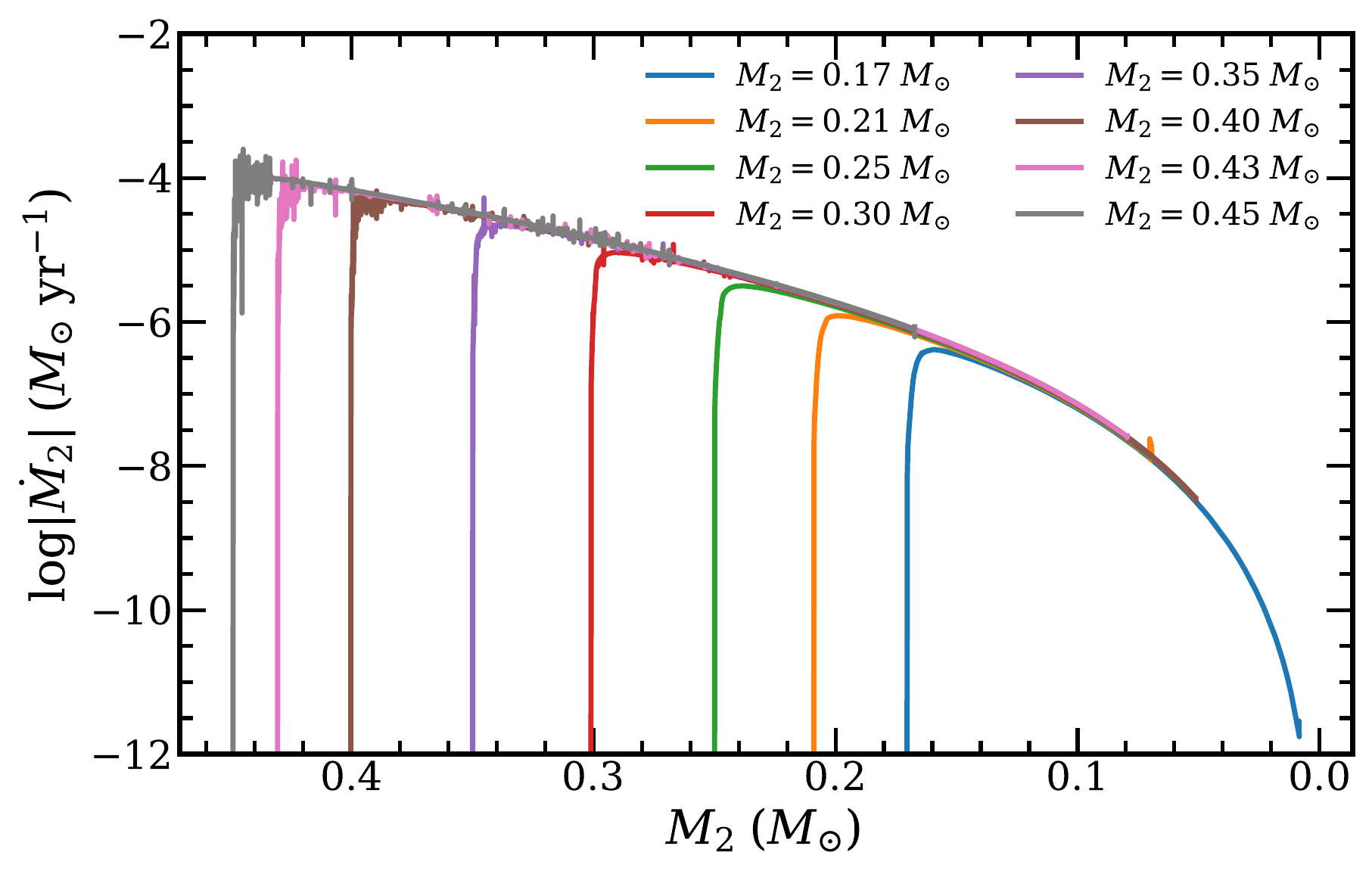}
    \includegraphics[width=\columnwidth]{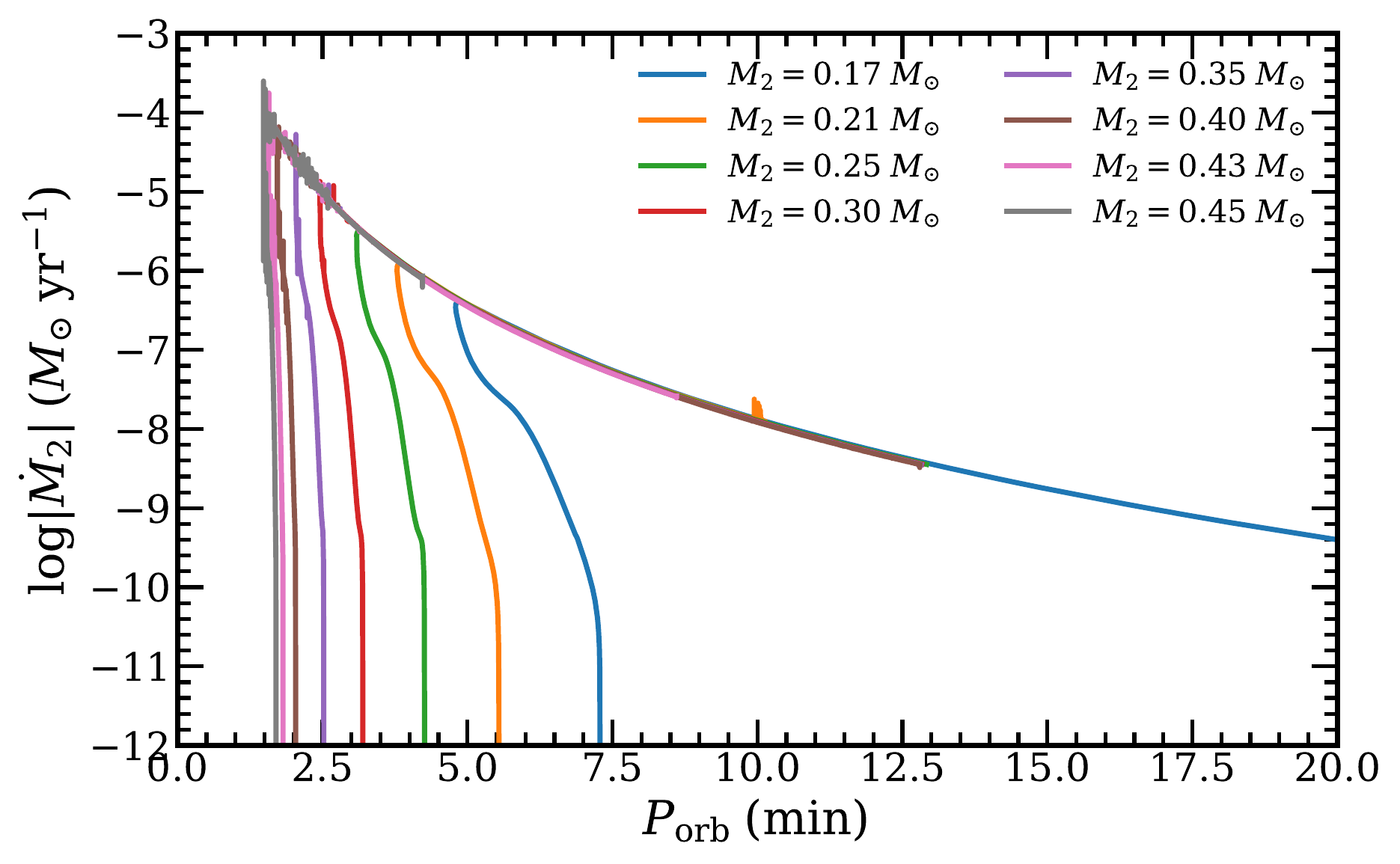}
    \caption{Evolution of mass-transfer rate as a function of WD mass (upper panel) and orbital period (lower panel) for NS+He~WD binaries with different WD masses. The initial orbital periods and WD properties are listed in Table~\ref{tab:ini_wd}.}
    \label{fig:bin_all}
\end{figure}

In Fig.~\ref{fig:bin_all}, we present the evolution of mass-transfer rate as a function of WD mass and orbital period for a number of NS+He~WD binaries with different WD masses. From this figure, we conclude that the maximum mass-transfer rate is larger for more massive WDs (as expected). Since massive WDs are more compact compared to lighter WDs, they initiate mass transfer at smaller orbital periods and reach smaller minimum orbital periods during RLO. The minimum orbital periods of these systems are between $\sim 1.5 - 4.8\;{\rm min}$, see Table~\ref{tab:ini_wd}.  

Our main result is that all NS+He~WD binaries evolve with stable mass transfer and survive the UCXB phase. 
This is consistent with the semi-analytic results of some previous studies \citep[e.g.][]{vr88,vnvt+13}, but the conclusion is very different from 
\citet{bdc17} who find that only systems with He~WD masses $<0.20\;M_{\odot}$ will undergo stable mass transfer in UCXBs by combining the results of hydrodynamic simulations with a semi-analytic long-term evolution model. This difference is likely due to the different assumptions of specific angular momentum loss in the different studies.  \citet{bdc17} (see their fig. 14) also found that the critical WD mass can be higher if they assumed that the mass loss from the systems takes away the specific angular momentum of the NS, as is the standard assumption in the isotropic re-emission model \citep{tv06} applied in MESA.

Another interesting point in Fig.~\ref{fig:bin_all} is that all the evolution tracks converge to a single branch after the mass-transfer rates reach their maximum, in agreement with the findings of \citet{stli17}.

\section{Discussion}\label{sec:dis}
\subsection{Influence of accretion disk}\label{subsec:dis_dis}
In our calculations, the effect of an accretion disk is not taken into account. 
\citet{vnvw+12} carefully discussed this point and found that as long as the mass ratio, $q=M_2/M_{\rm NS} \ga 0.01$, the torque between the outer disk and the WD donor star can efficiently transfer angular momentum back from the disk to the orbit. In this case, GW radiation and mass loss are the main orbital angular momentum loss mechanisms, which is consistent with our treatment. Only for $q \la 0.01$, the feedback of angular momentum from the disk to the orbit is inefficient and the disk becomes a net sink of angular momentum. This will lead to a sudden increase of mass-transfer rate \citep[cf. fig.~8 of][]{vnvw+12}. Thus we expect that our calculations at the very final stage, once $M_2 < 0.02\;M_\odot$, may not be reliable.
It should be noted that in Fig.~\ref{fig:bin_all}, only the model an initial WD mass of $0.17\;M_{\odot}$ was successfully evolved all the way to $0.008\;M_{\odot}$. 
The rest of the models terminated close to $\sim 0.05\;M_{\odot}$. However, it has been proposed that such
systems may anyway become tidally disrupted \citep{rs85} near this donor mass limit \citep[possibly leading to production of isolated radio MSPs with planets,][]{mlp16}.

\subsection{Influence of NS accretion efficiency}\label{subsec:acc_efficiency}
\begin{figure}
    \centering
    \includegraphics[width=\columnwidth]{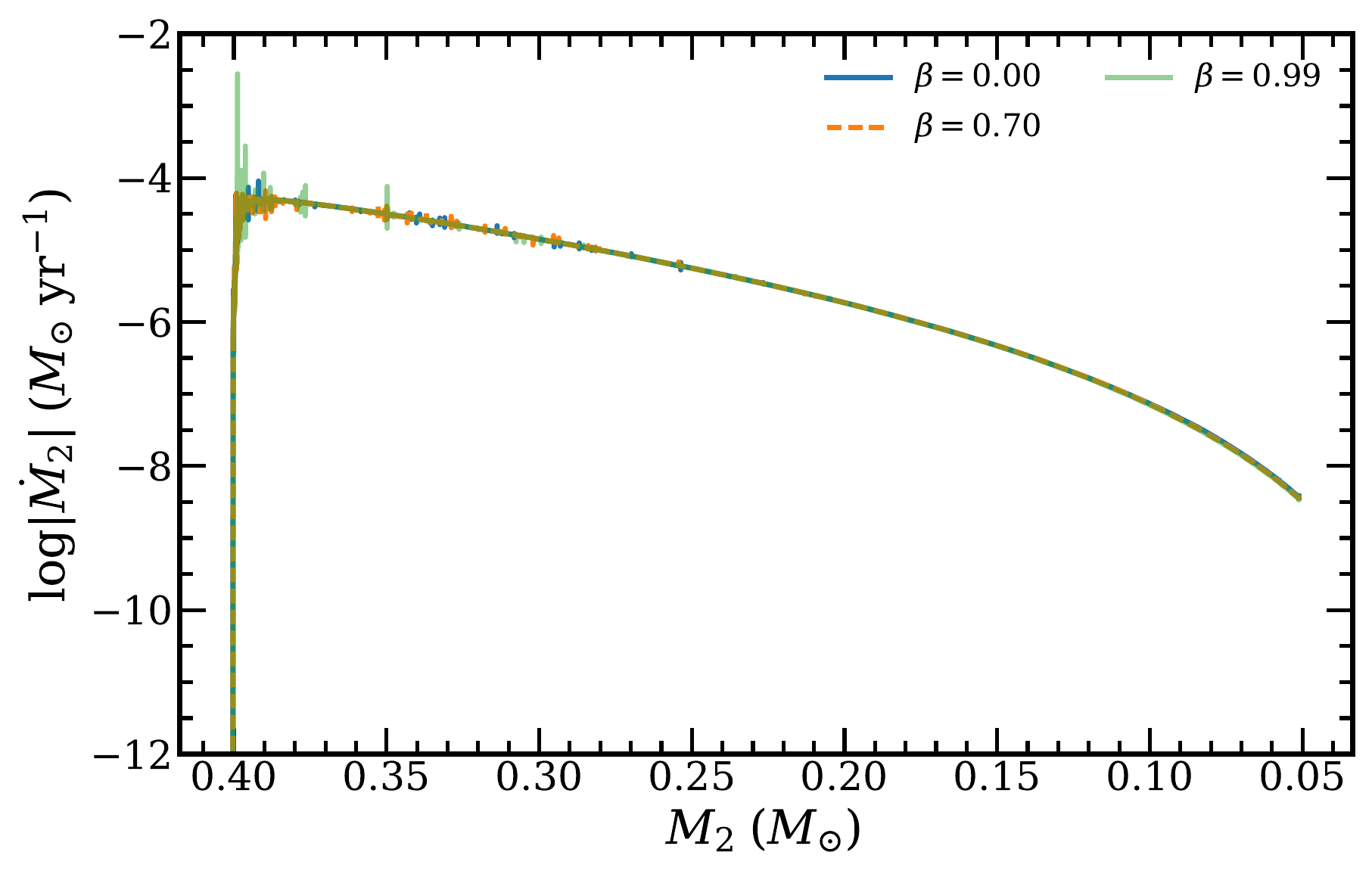}
    \caption{{\textbf Comparison of mass-transfer rates for NS+He~WD binaries with the same initial WD mass of $0.40\;M_{\odot}$, $M_{\rm NS}=1.30\;M_\odot$, and $P_{\rm orb}=0.02\;{\rm days}$ , but different $\beta$-values of 0.00, 0.70, and 0.99, respectively.}}
    \label{fig:acc_efficiency}
\end{figure}

The stability of our UCXB calculations are independent of the NS accretion efficiency. This is illustrated in Fig.~\ref{fig:acc_efficiency} where we have plotted three different computed tracks with $\beta =0.0$ (fully conservative RLO), 0.70 and 0.99 (almost complete re-ejection of all transferred material), respectively. In all cases the initial WD mass, NS mass, and orbital period, were assumed to be $0.40\;M_\odot$, $1.30\;M_\odot$ and $0.02\;{\rm days}$, respectively. 

\subsection{Influence of initial effective temperature and orbital period}\label{subsec:dis_pd}
Aside from isolated binary star evolution, NS+He~WD systems as UCXB sources may be produced from stable mass transfer after an episode of an exchange encounter event in a globular cluster, or they may form via the CE channel. Therefore, the effective temperature of the He~WD at the onset of RLO can take a range of different values. In our calculations, we applied roughly the same central temperatures for all the different WD donors. However, if the initial orbital period is long (short) after the formation of the NS+He~WD system (irrespective of its formation channel), the WD had long (short) time to cool down before initiating RLO. Thus also the effective temperature of the WD \citep[and thereby also its radius,][]{itla14} at the onset of mass transfer can take a range of values. Therefore, the effect of different initial effective temperatures should be similar to the effect of different initial orbital periods. 

\begin{figure}
    \centering
    \includegraphics[width=\columnwidth]{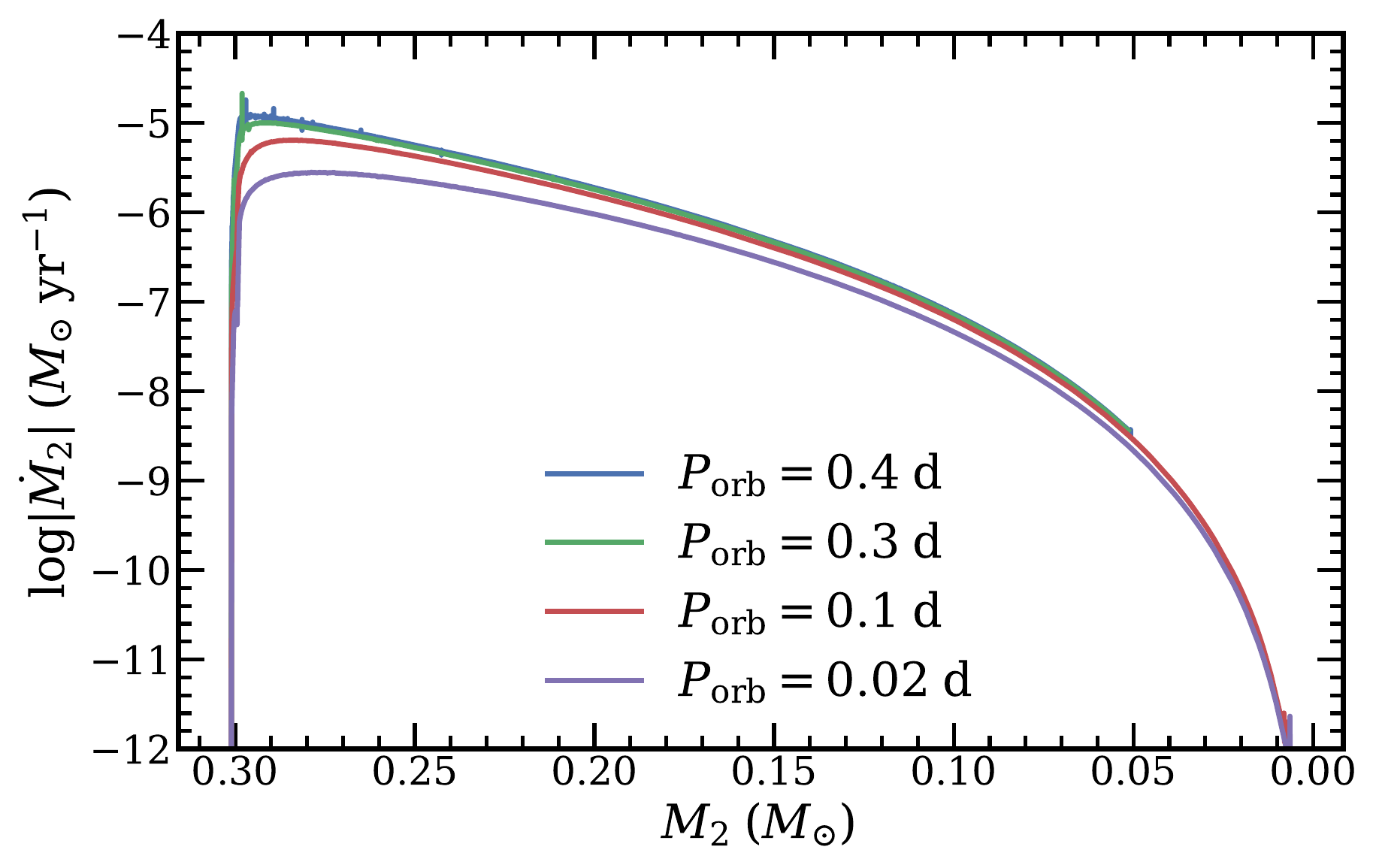}
    \caption{Comparison of mass-transfer rates for NS+He~WD binaries with the same initial WD mass of $0.30\;M_{\odot}$, but different initial orbital periods of 0.40, 0.30, 0.10, and $0.02\;{\rm days}$, respectively.}
    \label{fig:diff_orb}
\end{figure}

To better understand the influence of initial effective temperature, we produced a $0.30\;M_{\odot}$ He~WD with $T_{\rm eff} = 3.85 \times 10^{4}\;{\rm K}$ (and a central temperature ${\rm log}(T_{\rm c}/{\rm K}) = 7.64$) using the method described in Section~\ref{sec:met}. We then compared the evolution of mass-transfer rate for NS+He~WD binaries with different initial orbital periods. Our results are shown in Fig.~\ref{fig:diff_orb}. 
We find that the mass-transfer rates are somewhat higher (up to one order of magnitude) for binaries with larger initial orbital periods. This is because these systems have longer WD cooling times and therefore become more degenerate. Except for this dissimilarity around the peak of mass-transfer rate, the difference between these binaries with different orbital periods is much less pronounced as the systems evolve further. We restricted our computations to binaries with initial orbital periods short enough to initiate the UCXB stage within the age of the Universe. As an example, the model with an initial orbital period of $0.40\;{\rm days}$ takes $12.26\;{\rm Gyr}$ to become semi-detached. On top of this should be added the time of the previous evolution before producing the NS+He~WD binary, which may also take several Gyr. Therefore, we do not consider binaries with longer orbital periods here.

\subsection{GW sources}
\begin{figure}
    \centering
    \includegraphics[width=\columnwidth]{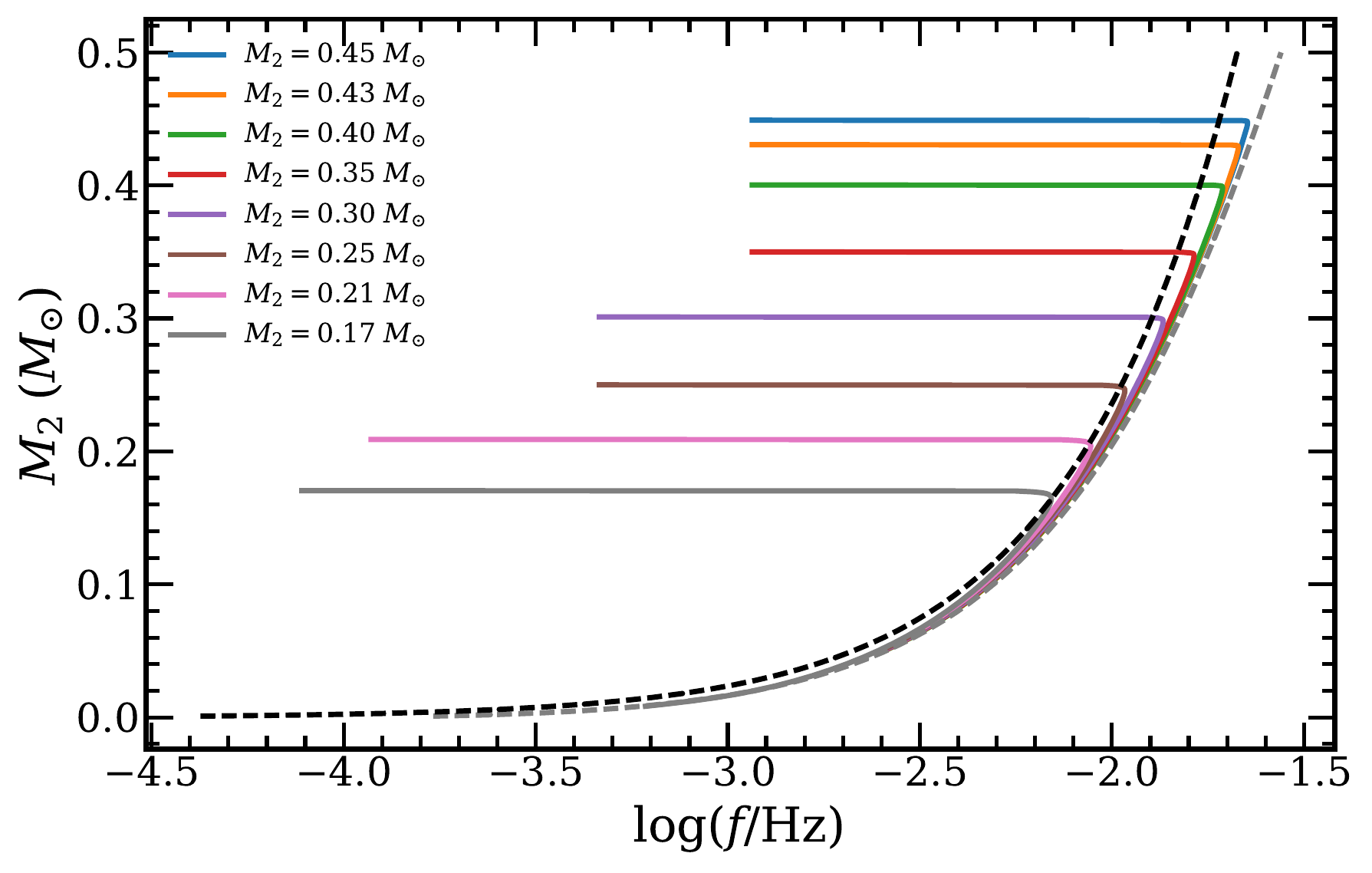}
    \caption{Evolution of donor mass as a function of GW frequency for NS+He~WD binaries with different WD masses. The black and grey dashed lines indicate the analytic results of Eqs.~(\ref{eq:fwd1}) and (\ref{eq:fwd2}), respectively. }
    \label{fig:fgw_mwd}
\end{figure}

Given the short orbital periods of UCXBs, they are strong GW emitters in the frequency band $1-10\;{\rm mHz}$. Therefore, they are important GW sources for space-borne GW observatories like LISA \citep{aabb+17}, TianQin \citep{lcdg+16}, Taiji \citep{rgcz20}.  

In Fig.~\ref{fig:fgw_mwd}, we show the evolution of WD mass as a function of GW frequency. It is clearly seen that all evolutionary tracks converge to the same branch after the minimum orbital periods (maximum peak in frequencies) are reached. Therefore, there is a tight relation between the GW frequency and the WD mass for UCXBs detected in GWs. 
This relation can be understood as outlined in the following.

According to the Kepler's third law, the GW frequency (twice the orbital frequency) is given by:
\begin{equation}
    f = \frac{2}{P_{\rm orb}} = \frac{1}{\pi} \sqrt{\frac{G(M_{\rm  NS}+M_{2})}{a^{3}}}\;,
\label{eq:fgw}
\end{equation}
where $P_{\rm orb}$ is the orbital period; $M_{\rm NS}$ and $M_{2}$ are the NS mass and He WD mass, respectively; and $a$ is the binary separation. 
The Roche-lobe radius of the He~WD can be estimated with the following equation \citep{pacz1971}:
\begin{equation}
    \frac{R_{\rm rl}}{a} \simeq 0.462\,\left(\frac{M_2}{M_{\rm NS}+M_2}\right)^{1/3}\;,
\label{eq:rl}
\end{equation}
which is a good approximation for $M_2/M_{\rm NS}<0.523$ (and therefore all UCXBs), and where $R_{\rm rl}$ is the Roche-lobe radius of the He~WD.
During the mass-transfer phase, the radius of He~WD ($R_{2}$) equals that of its Roche lobe:
\begin{equation}
    R_2 = R_{\rm rl}\;.
\label{eq:rrl}
\end{equation}
For a relation between WD mass and radius, we may at first apply \citep{chan39}:
\begin{equation}
R_{2} = 0.013\;R_{\odot}\left(\frac{M_2}{M_\odot}\right)^{-1/3}\;.
\label{eq:rmwd1}
\end{equation}
Combining Eqs.~(\ref{eq:fgw})--(\ref{eq:rmwd1}), we obtain:
\begin{equation}
  f=C_1\,M_2 \;,
\end{equation}
where the constant $C_1$ is given by:
\begin{equation}
  C_1=\frac{1}{\pi}\sqrt{\frac{(0.462)^3\,G}{(0.013\;R_\odot)^3M_\odot}}
  = 2.13\times 10^{-35}\;{\rm Hz\,g}^{-1}\;.
\end{equation}
Rewriting in more convenient units, we finally obtain:
\begin{equation}
  f=4.23\;{\rm mHz}\;\left(\frac{M_2}{0.1\;M_\odot}\right)
\label{eq:fwd1}
\end{equation}
It is interesting to note that the GW frequency, $f$ is {\em independent} of the accreting NS mass and only depends on the He~WD mass, $M_2$.

Alternatively, applying a slightly more accurate WD mass--radius relation, first proposed by Peter Eggleton and quoted in \citet{vr88}:
\begin{eqnarray}
  R_2 & = & 0.0114\;R_\odot\left[\left(\frac{M_2}{M_{\rm Ch}}\right)^{-2/3}-\left(\frac{M_2}{M_{\rm Ch}}\right)^{2/3}\right]^{1/2} \nonumber \\
 & &\times \left[1+3.5\left(\frac{M_2}{M_p}\right)^{-2/3}+\left(\frac{M_2}{M_p}\right)^{-1}\right]^{-2/3}\;,
\label{eq:rmwd2}
\end{eqnarray}
where $M_{\rm Ch}\simeq 1.44\;M_\odot$ is the Chandrasekhar mass limit for a WD and $M_p=0.00057\;M_{\odot}$ is a constant, we find:
\begin{eqnarray}
  f & = & C_2\,M_2^{1/2}\,\times \left[\left(\frac{M_{2}}{M_{\mathrm{Ch}}}\right)^{-2 / 3}-\left(\frac{M_{2}}{M_{\mathrm{Ch}}}\right)^{2 / 3}\right]^{-3/4}\nonumber \\
  & & \times \left[1+3.5\left(\frac{M_{2}}{M_{p}}\right)^{-2 / 3}+\left(\frac{M_{2}}{M_{p}}\right)^{-1}\right]
\end{eqnarray}
where the constant $C_2$ is given by:
\begin{equation}
  C_2=\frac{1}{\pi}\sqrt{\frac{(0.462)^3\,G}{(0.0114\;R_\odot)^3}}
  = 1.16\times 10^{-18}\;{\rm Hz\,g}^{-1/2}\;,
\end{equation}
or rewritten in more convenient units:
\begin{eqnarray}
  f & = & 16.3\;{\rm mHz}\,\left(\frac{M_2}{0.1\;M_\odot}\right)^{1/2}\nonumber \\
  & & \times \left[\left(\frac{M_2}{M_{\rm Ch}}\right)^{-2/3}-\left(\frac{M_2}{M_{\rm Ch}}\right)^{2/3}\right]^{-3/4}\nonumber \\
  & & \times \left[1+3.5\left(\frac{M_2}{M_p}\right)^{-2/3}+\left(\frac{M_2}{M_p}\right)^{-1}\right]
\label{eq:fwd2}
\end{eqnarray}
Considering that the bulk of detectable LISA sources are found where the detector has its highest sensitivity (near 5~mHz), we can rewrite the expression in the limit where $M_2\sim 0.1\;M_\odot$ and we obtain here:
\begin{equation}
  f\simeq 4.89\;{\rm mHz}\;\left(\frac{M_2}{0.1\;M_\odot}\right)^{1/2}
\label{eq:fwd2_approx}
\end{equation}

In Fig.~\ref{fig:fgw_mwd}, we have plotted the expressions from Eqs.~(\ref{eq:fwd1}) and (\ref{eq:fwd2}) as dashed lines and they are found to be quite consistent (especially Eq.~(\ref{eq:fwd2})) with the models from the \textsc{mesa} calculations. 
As a bonus of the independence of the frequency on NS mass, the WD mass can be directly inferred from observations of UCXBs by only measuring the GW frequency.

\subsection{Influence of initial NS mass}\label{subsec:initial_NS_mass}
\begin{figure}
    \centering
    \includegraphics[width=\columnwidth]{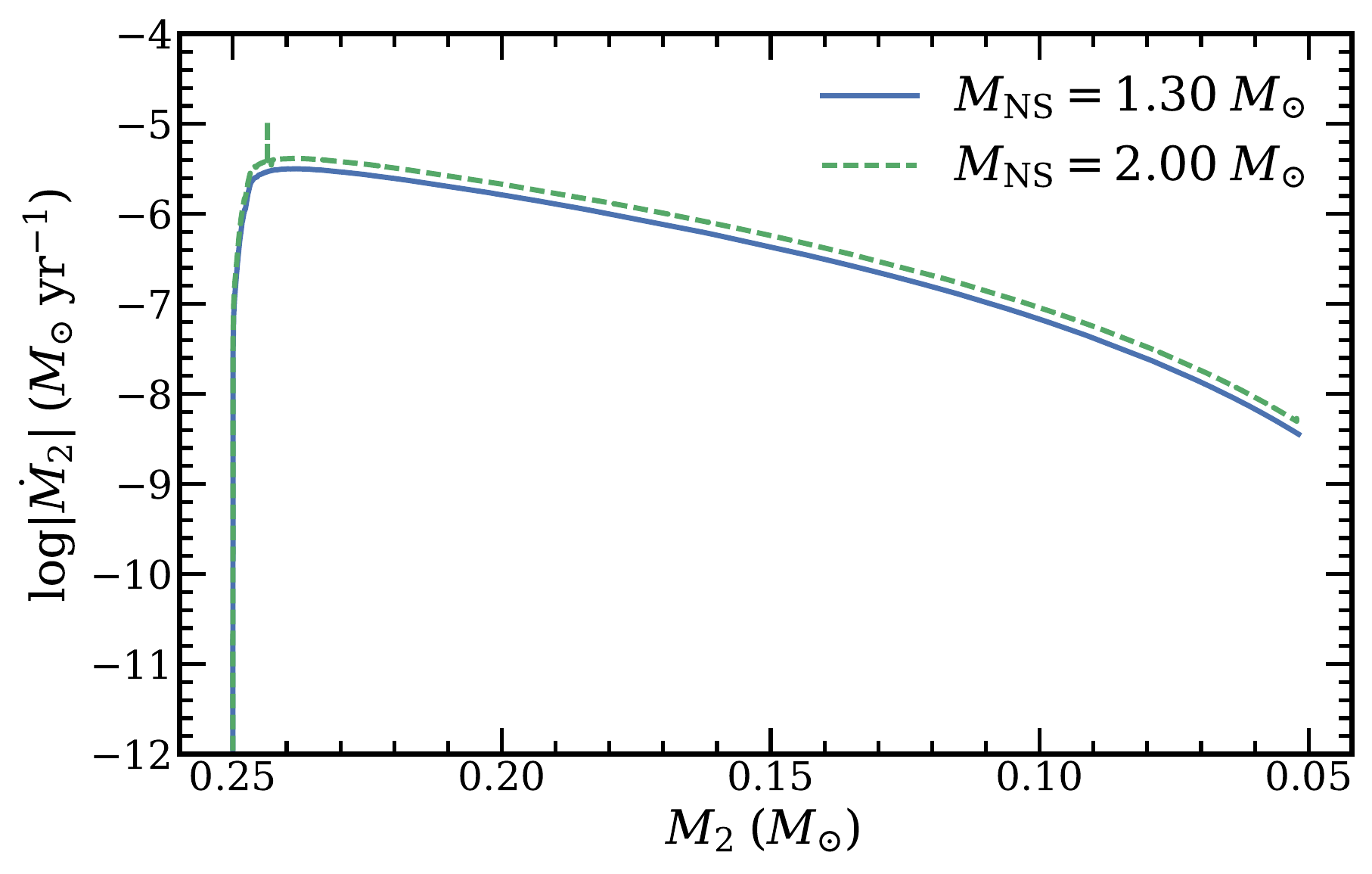}
    \includegraphics[width=\columnwidth]{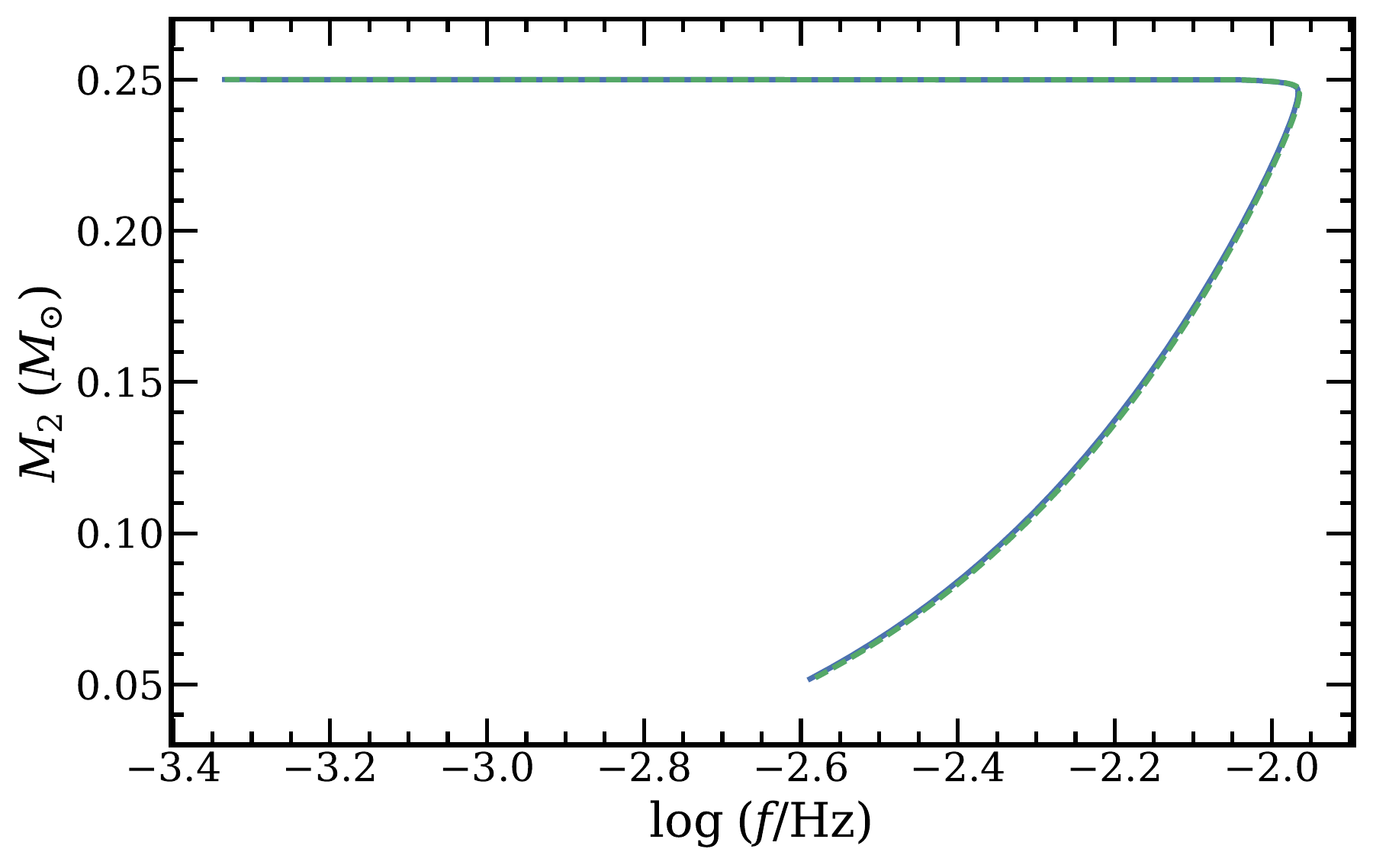}
    \caption{Comparison of the evolution of NS+He~WD UCXBs with different initial NS masses. The initial donor star masses (and temperatures) and orbital periods of these two models are the same ($0.25\;M_{\odot}$ and $0.05\;{\rm days}$, respectively). 
    The upper panel shows the evolution of mass-transfer rate as a function of decreasing donor mass. The lower panel shows the evolution of donor WD mass as a function of GW frequency. The blue solid and green dashed lines are for the two models with an initial NS mass of $1.30\;M_{\odot}$ and $2.0\;M_{\odot}$, respectively. 
    }
    \label{fig:diff_nsm}
\end{figure}

To further investigate the influence of the NS mass on the general UCXB evolution, we computed a NS+He~WD model with an initial NS mass of $2.0\;M_{\odot}$, an initial He~WD mass of $0.30\;M_{\odot}$ and an initial orbital period of $0.05\;{\rm days}$. In Fig.~\ref{fig:diff_nsm}, we compare this model directly to a similar model with an initial NS mass of $1.30\;M_{\odot}$. From the two panels of this figure, we conclude that the influence of initial NS mass on the general UCXB evolution is very limited.

\subsection{Hydrogen abundance in UCXBs}
\begin{figure}
    \centering
    \includegraphics[width=\columnwidth]{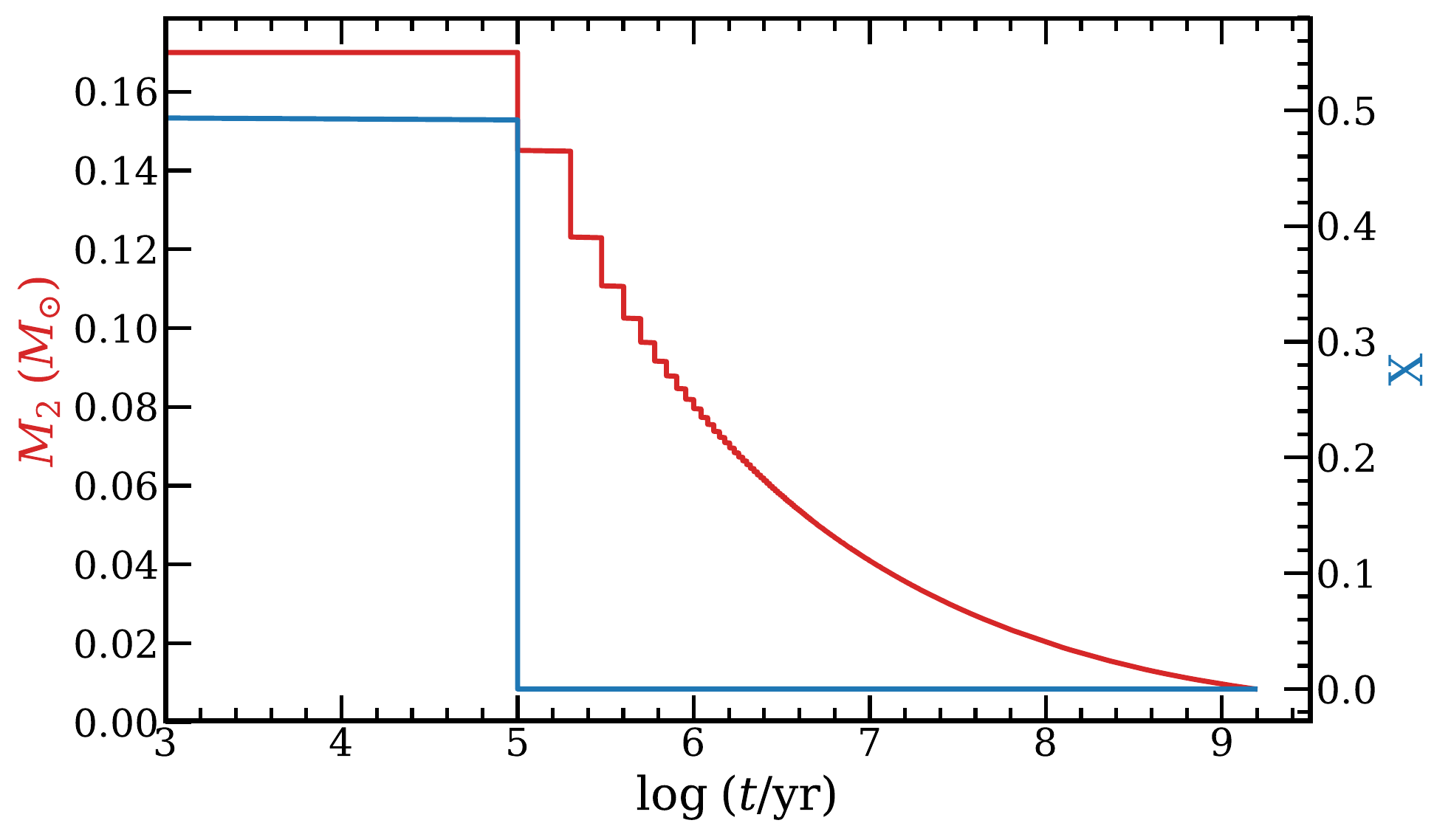}
    \includegraphics[width=\columnwidth]{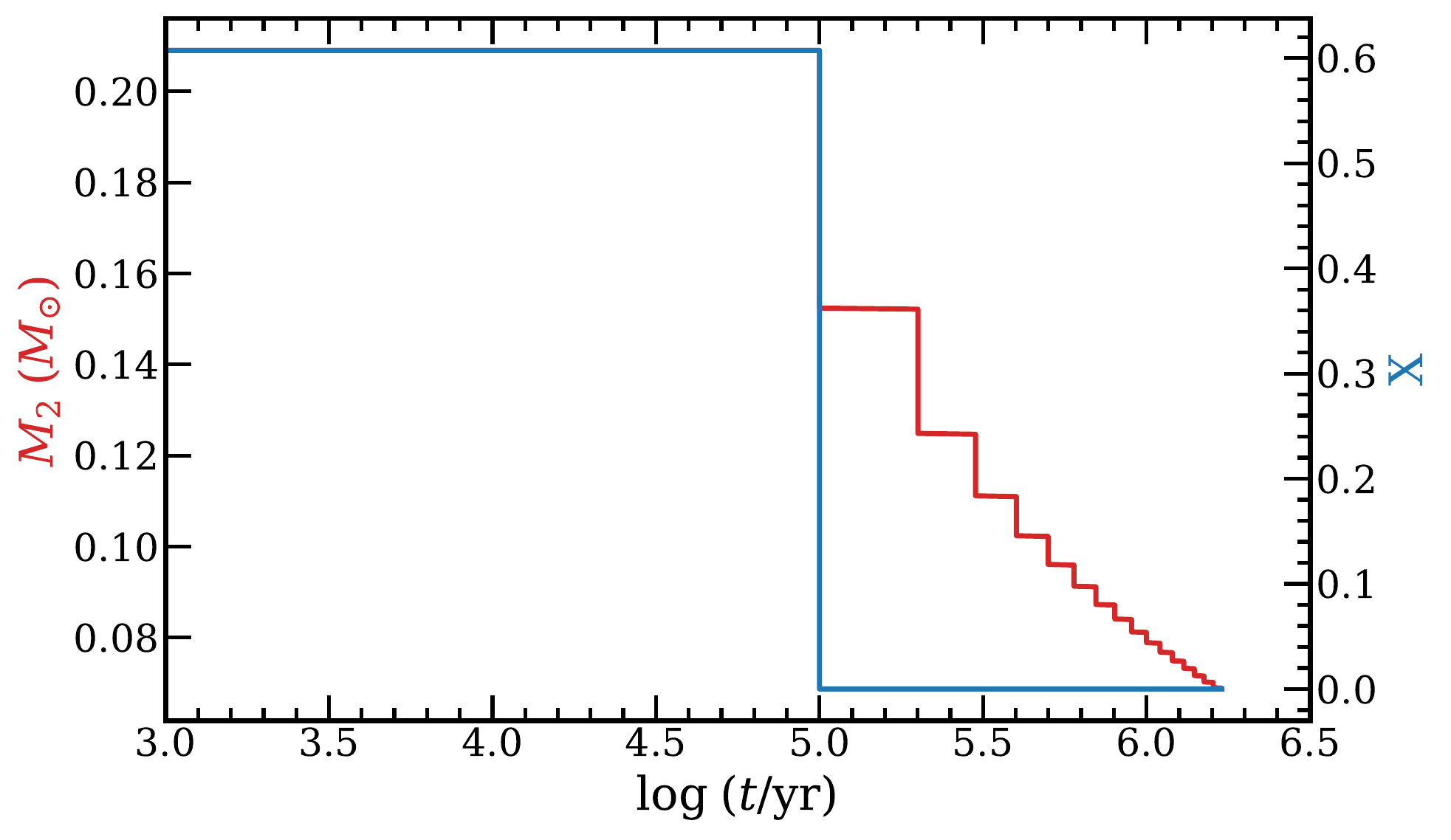}
    \includegraphics[width=\columnwidth]{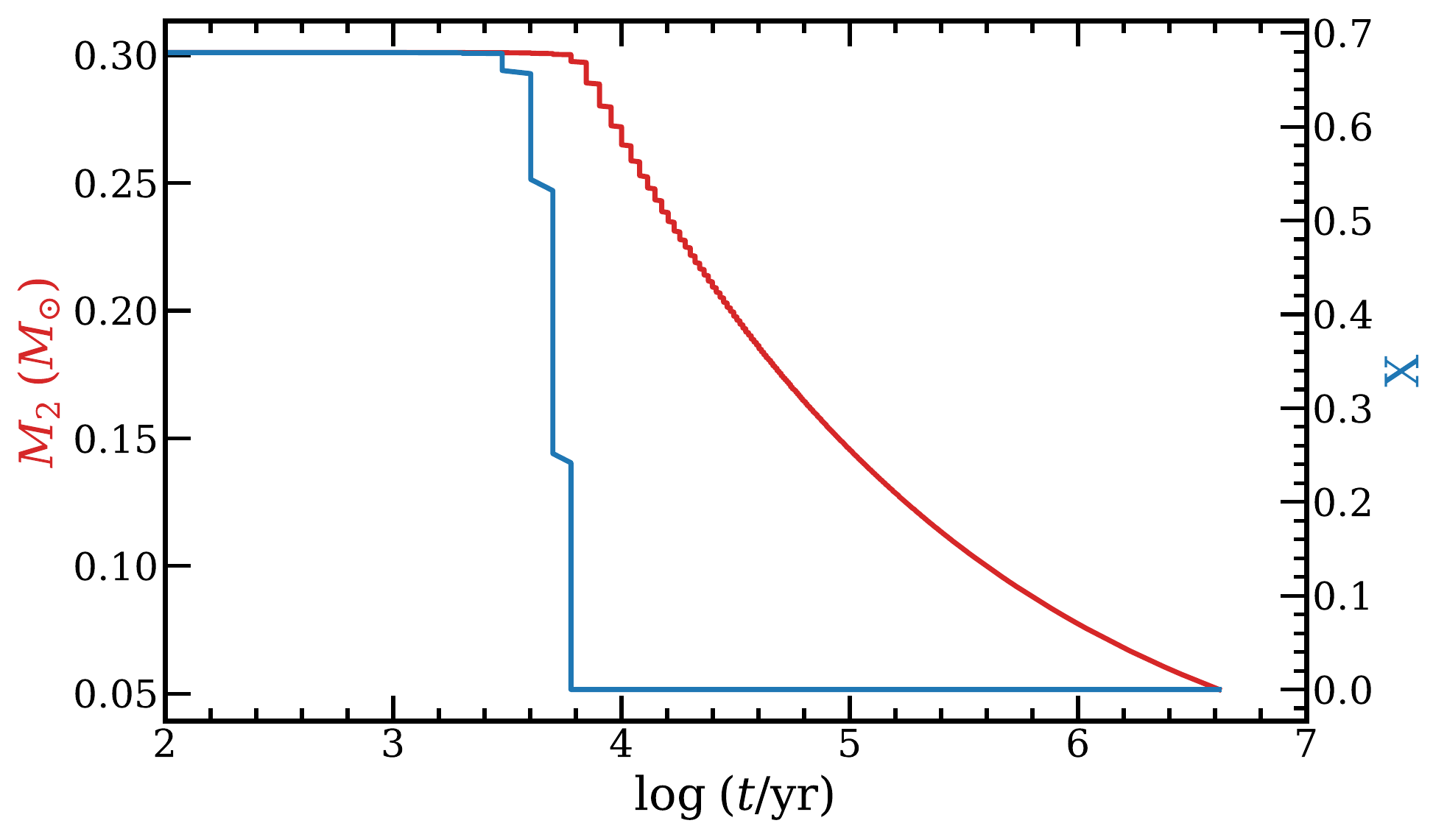}
    \includegraphics[width=\columnwidth]{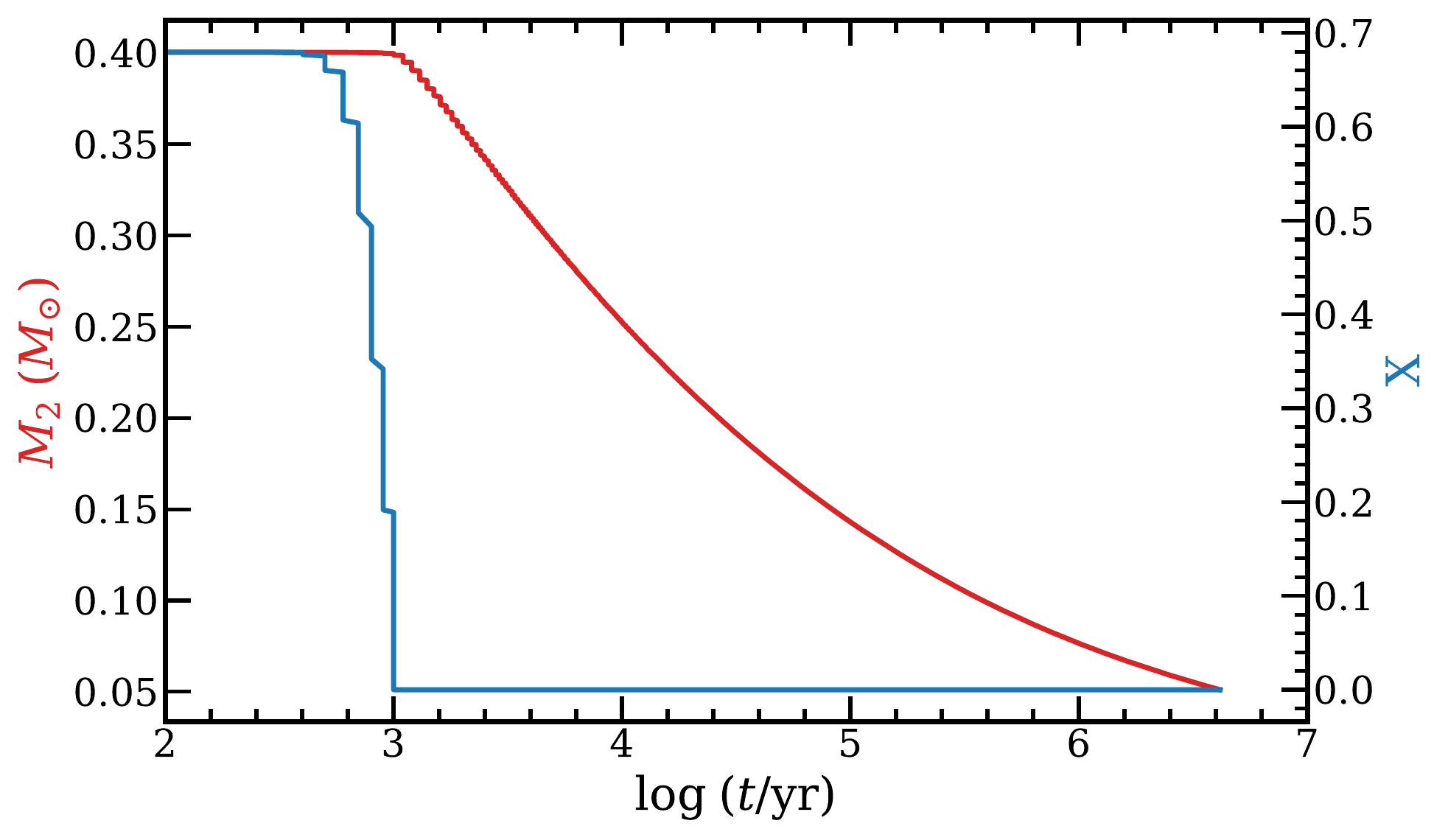}
    \caption{ Evolution of surface H abundance (blue lines) and WD mass (red lines) as a function of time for UCXBs with different initial WD masses. From top to bottom, the initial WD masses are $0.17$, $0.21$, $0.30$, $0.40\;M_{\odot}$, respectively.  The $t = 0$ point indicates the onset of mass transfer ($\mathrm{log}\;({\dot{M_{2}}/(M_{\odot}/{\mathrm{yr}})}) >= -12.0$). The left and right y-axes indicate WD mass and surface H abundance, respectively.}
\label{fig:md_x_evl}
\end{figure}

Given that He~WDs are produced with an outer layer of residual hydrogen (H) from the LMXB stage, we expect that H could be present in the accretion disc of UCXBs in their early phases. In Fig.~\ref{fig:md_x_evl}, we present the evolution of surface H abundance and mass of the He~WD donors as a function of time for models with initial He~WD masses of $0.17$, $0.21$, $0.30$, $0.40\;M_{\odot}$. From this plot, we see that H is only expected to be present in the accretion disc of an UCXB at its very early stage of the mass transfer, lasting for about $10^{3} - 10^{5}\;{\rm yr}$. Given this short time interval, it will be difficult to observe the presence of H in the accretion disc.

\section{Conclusion}\label{sec:con}
With the stellar evolution code \textsc{mesa}, we modelled the evolution of NS+He WD binaries with WD masses ranging from $0.17\;M_{\odot}$ to $0.45\;M_{\odot}$. The main results are as follows:

\begin{itemize}
    \item All NS+He~WD binaries (with an initial WD mass anywhere in the interval between $\sim 0.17-0.45\;M_\odot$) undergo {\em stable} UCXB mass transfer. This is a prediction that is, in principle, easily testable with LISA, TianQin and Taiji GW observations, although population synthesis is needed to reveal how many systems are detectable, given that the WD mass initially decreases rapidly.
    \item The larger the WD masses, the larger the maximum mass-transfer rates, and the smaller the minimum orbital periods of the NS+He~WD systems. The minimum orbital periods range between $1.5\;{\rm min}$ to $4.7\;{\rm min}$, corresponding to maximum GW frequencies of $f\approx 7.1-22\;{\rm mHz}$. 
    \item There is a tight correlation between WD mass and GW frequency for NS+He~WD UCXBs, {\em independent} of NS mass. With the relation in Eq.(\ref{eq:fwd2}) (or Eq.~\ref{eq:fwd1}), the WD mass can be inferred in all such UCXBs for which the GW frequency can be determined from observations. A measurement of the chirp mass, $\mathcal{M}(f,\dot{f})\equiv (M_1 M_2)^{3/5}/(M_1 + M_2)^{1/5}$, will thus directly provide the NS mass.
    \item It is very unlikely that hydrogen will be detected in the accretion disk of UCXBs with a He~WD donor star because their residual outer layer of hydrogen ($\la 0.01\;M_\odot$, left from the detached LMXB progenitor system) is removed within $10^5\;{\rm yr}$.
\end{itemize}

\begin{acknowledgments}
We thank the anonymous referee for comments that helped improving the manuscript.
This work is partially supported by the National Natural Science Foundation of China (Grant No. 12090040/12090043, 2021YFA1600400/401, 12073071, 11873016, 11733008), 
Yunnan Fundamental Research Projects (Grant No. 202001AT070058, 202101AW070003), the science research grants from the China Manned Space Project with No. CMS-CSST-2021-A10) and Youth Innovation Promotion Association of Chinese Academy of Sciences (Grant no. 2018076). The authors gratefully acknowledge the ``PHOENIX Supercomputing Platform'' jointly operated by the Binary Population Synthesis Group and the Stellar Astrophysics Group at Yunnan Observatories, CAS.
We are grateful to the \textsc{mesa} council for the \textsc{mesa} instrument papers and website. 
\end{acknowledgments}

\vspace{5mm}
\software{MESA \citep{pbdh+11,pcab+13,pmsb+15,psbb+18,pssg+19}  
          }

\bibliography{hailiang_refs.bib}
\bibliographystyle{aasjournal}

\end{document}